
%
%
\input harvmac
\Title{\vbox{\baselineskip14pt\hbox{IMSc--92/36}\hbox{hepth@xxx/9209051}}}{
On Physical States in $c<1$ String Theory}
\centerline{\titlefont {On Physical States in $c<1$ String Theory}}
\bigskip
\centerline{Suresh Govindarajan\footnote{$^\dagger$}{Talk presented at the
Worshop on Superstrings and Related Topics at ICTP, Trieste in July 1992.}}
\centerline{The Institute of Mathematical Sciences}
\centerline{C.I.T. Campus, Taramani}
\centerline{Madras 600 113, INDIA}
\bigskip
\bigskip
\noindent
{\bf Abstract:~}The BRST cohomology analysis of Lian and Zuckerman
leads to physical states at all ghost number for $c<1$ matter coupled
to Liouville gravity. We show how these states are related to states at ghost
numbers zero(pure vertex operator states -- DK states) and ghost
number one(ring elements) by means of descent equations. These descent
equations follow from the double cohomology of the String BRST and
Felder BRST operators. We briefly discuss how the ring elements allow
one to determine all correlation functions on the sphere.
\bigskip
\Date{September 1992}
\lref\SWA{Ghoshal, D.  and Mahapatra, S., ``Three-Point Functions of
Non-Unitary Minimal Matter coupled to gravity,'' Tata Preprint
TIFR/TH/91-58(1991).}
\lref\NSYT{Sakai, N. and Tanii, Y.,
           Int.\ J. of Mod.\ Phys.\ {\underbar {A6}}, 2743(1991).}
\lref\DF{Dotsenko, Vl. and Fateev, \npb\ {\underbar {B251}}, 691(1985).}
\lref\SAR{H. Kanno and H. Sarmadi, Talk presented by H. Sarmadi at the
Summer Workshop on Strings at ICTP, Trieste(July 1992).}
\lref\SARa{H. Kanno and H. Sarmadi, ``BRST cohomology ring in 2D gravity
coupled to minimal models,'' ICTP preprint IC/92/150 =
hepth@xxx/9207078}
\lref\DOY{M. Doyle, ``Dilaton Contact Terms in the Bosonic and Heterotic
Strings,'' Princeton preprint PUPT-1296(1992) =
hepth@xxx/9201076.}
\lref\BT{R. Bott and L. W. Tu, {\it Differential Forms in Algebraic
Topology}, Springer-Verlag(1982).}
\lref\KACH{S. Kachru, ``Quantum Rings and Recursion Relations in 2D Quantum
Gravity,'' Princeton Preprint PUPT-1305 = hepth@xxx/9201072.}
\lref\ITOH{K. Itoh, ``SL(2,R) current algebra and spectrum in two
dimensional gravity,'' Texas A\&M preprint CTP-TAMU-42/91(1991).}
\lref\AGG{L. Alvarez-Gaum\'e and C. Gomez, ``Topics in Liouville Theory,''
Lectures at Trieste Spring School, CERN preprint CERN-TH.6175/91, (1991).}
\lref\GRO{U. H. Danielsson and D. Gross, \npb{\bf 366}(1991), 3.}
\lref\FKa{P. Di Francesco and D. Kustasov, \npb{\bf 342}(1990), 589.}
\lref\FKb{P. Di Francesco and D. Kutasov, ``World Sheet and Space time
Physics in two dimensional (super) string theory,'' Princeton preprint
PUPT-1276(1991).}
\lref\DOT{Dotsenko, V., \mpl{\underbar {A6}}, 3601 (1991).}
\lref\GF{Felder, G., Nuc. Phys. {\underbar {B317}}, 215(1989).}
\lref\LZ{Lian, B. and Zuckerman, G., Phys. Lett.  {\underbar {B254}},
417 (1991);  \cmp{\underbar {145}}, 561(1992).}
\lref\IMM{C. Imbimbo, S. Mahapatra and S. Mukhi,``Construction of Physical
States of Non-trivial ghost number in $c<1$ String Theory,'' \npb375(1992),
399.}
\lref\GLI{Goulian, M. and  Li, B.,\prl{\underbar {66}}, 2051(1991).}
\lref\BMP{ Bouwknegt, P., McCarthy, J. and  Pilch, K.,
 \cmp{\underbar {145}}, 541(1992).}
\lref\BMPa{P. Bouwknegt, J. McCarthy and K. Pilch, ``Fock Space Resolutions
of the Virasoro highest weight modules with $c\leq1$,'' Lett. Math. Phys.
23(1991), 3601.}
\lref\KIT{Kitazawa, Y., \plb {\underbar {B265}}, 262(1991).}
\lref\KITa{Y. Kitazawa, ``Puncture Equation in $c=1$ Liouville
gravity,'' TIT preprint TIT(1991).}
\lref\KM{M. Kato and S. Matsuda in {\it Advanced Studies in Pure Mathematics},
Vol. 16, ed. H. Morikawa(1988), 205.}
\lref\POL{A. M. Polyakov, \mpl{\bf 6}(1991), 635-644.}
\lref\FF{B. Feigin and D. Fuchs, ``Representations of the Virasoro algebra,''
in {\it Seminar on Supermanifolds} No.5, ed. D. Leites(1988), Univ. of
Stockholm Report No. 25.}
\lref\JD{J. Distler, \npb{\bf 342}(1990), 523.}
\lref\DN{J. Distler and P. Nelson, ``New discrete states of strings
near a black hole,'' Penn preprint UPR-0462T(1991).}
\lref\KAL{S. Kalyana Rama, ``New special operators in W-gravity theories,''
Tata preprint TIFR/TH/91-41(1991).}
\lref\SEI{N. Seiberg,``Notes on Quantum Liouville Theory and Quantum Gravity,
,'' Prog. of Theo. Phys., {\bf 102}(1990), 319.}
\lref\POLC{J.Polchinski, ``Remarks on the Liouville field theory,'' Texas
preprint UTTG-19-90, in Proceedings of Strings '90.}
\lref\BER{Bershadsky, M. and Klebanov, I., \prl\ {\underbar {65}}, 3088(1990);
\npb\ {\underbar {B360}}, 559(1991).}
\lref\KAED{K. Aoki and E. D'Hoker, ``On the \lio\ approach to correlation
functions for 2-D quantum gravity,'' UCLA preprint UCLA/91/TEP/32(1991).}
\lref\descent{Govindarajan, S., Jayaraman, T., John, V. and Majumdar, P.
Mod. Phys. Lett. {\underbar {A7}}, 1063(1992).}
\lref\DOTa{Vl. M. Dotsenko, `` Remarks on the physical states and the
chiral algebra of 2d gravity coupled to $c\leq1$ matter,''
PAR-LPTHE 92-4(1992) = hepth@xxx/9201077;  ``The operator algebra
of the discrete state operators in 2D gravity with
non-vanishing cosmological constant,''
CERN-TH.6502/92 =
PAR-LPTHE 92-17.}
\lref\EW{Witten, E., \npb{\underbar {B373}}, 187(1992).}
\lref\KMS{ Kutasov, D.,  Martinec, E. and Seiberg, N., \plb{\underbar {B276}},
437(1992).}
\lref\KLEB{I. R. Klebanov, ``Ward Identities in 2d gravity,''
Mod. Phys. Lett. {\bf A7}, (1992)723,}
\lref\CDK{N. Chair, V. Dobrev and H. Kanno, ``$SO(2,{\bf C})$
invariant ring structure of BRST cohomology and singular vectors in 2d
gravity with $c<1$ matter,'' \plb{\bf 283}(1992), 194.}
\lref\EWBZ{E. Witten and B. Zweibach, ``Algebraic structures and
Differential Geometry in 2d string theory,'' \npb{\bf 377}(1992), 55.}
\lref\JDTG{J. Distler, ``2-d quantum gravity, topological field theory,
and the multicritical matrix models,'' Nucl. Phys. {\bf B342} (1990) 523.}%
\lref\TCCT{J. Distler and P. Nelson, ``Topological couplings and
contact terms in 2d field theory,''
Commun. Math. Phys. {\underbar {138}}, 273 (1991).}%
\lref\dil {J. Distler and P. Nelson, ``The dilaton
equation in semirigid string theory,''  PUPT-1232 = UPR0428T (1991),
Nucl. Phys. {\bf B}, in press.}
\lref\ring{Govindarajan, S., Jayaraman, T. and John, V., ``Chiral Rings
and Physical States in $c<1$ String Theory,'' Preprint
IMSc--92/30=hepth@xxx/9207109.}
\lref\corr{Govindarajan, S., Jayaraman, T. and John, V., ``Genus Zero
Correlation Functions in $c<1$ String Theory,'' Preprint
IMSc--92/35=hepth@xxx/9208064.}
\lref\GJJ {S.Govindarajan, T. Jayaraman and V. John, Matscience preprint
IMSc--92/31, to appear}
\lref\mukhi{Mukhi, S., ``An introduction to continuum non-critical
strings,'' Preprint TIFR/ TH/92--05.}

\def\plb{Phys. Lett. }
\def\prl{Phys. Rev. Lett.}
\def\mpl{Mod. Phys. Lett. }
\def\npb{Nuclear Phys. }
\def\cmp{Comm. Math. Phys.}
\def\ket#1{| #1 \rangle}
\def\melt#1#2#3{\langle #1 \mid #2 \mid #3 \rangle}
\def\lfr#1#2{\textstyle{#1 \over #2 }}
\def\al{{\alpha}}
\def\pa{{\partial}}

\def\lio{Liouville}

\def\um#1#2{\ket{u_{{#1},{#2}}}_M}

\def\gh{c_1\ket0_{gh}}
\def\dot#1#2{\ket{\al_{{#1},{#2}}}\otimes\ket{\beta}\otimes\gh}

\newsec{INTRODUCTION}
Non-critical string theory received a boost when the discretised model
-- the matrix models were exactly solved to all orders in string
perturbation theory. The continuum limit of the matrix models --
Liouville theory has not been understood to the same extent as matrix
models. Early progress was made in calculating the torus partition
function$^{\BER\NSYT}$ followed by the calculation of three point functions
on the sphere$^{\GLI\DOT\KIT\SWA}$. The BRST analysis  of Lian
and Zuckerman$^{\LZ}$
and subsequently by BMP$^{\BMP}$ has shown that physical states
for $c<1$ string occur at all ghost numbers. This is unlike in the
critical bosonic string where it is restricted to three ghost numbers.
In this talk, I shall describe how the physical states in $c<1$
string theory can be represented by states at three ghost numbers just as in
the critical bosonic string. We shall work in the free field
formulation of the minimal models. The truncation of the Hilbert space
of a single scalar field to that of the finite space of primaries of
the minimal models is easily seen due to the existence of a BRST
operator -- the Felder BRST$^{\GF}$. On coupling to gravity, one also
has to deal with the usual String BRST operator. The double cohomology
of the the two BRST operators provides descent equations which relate
the LZ states of arbitrary ghost number to those of ghost numbers
similar to those occuring in the usual critical bosonic string. In
this talk, I will be discussing work appearing in $^{\descent\ring}$.
I shall also describe new results which have enabled computation of
all correlation functions on the sphere$^{\corr}$. These results agree
with the results obtained in matrix models.

\newsec{MINIMAL MODELS}

It is well known that a single scalar field with a background charge
provides a Fock space realisation of all minimal models. We shall
briefly explain how this works(see Dotsenko-Fateev$^{\DF}$ and Felder$^{\GF}$
for more details). Consider a scalar field with the following energy-momentum
tensor
\eqn\estressa{
T_M = -\lfr14 \pa X \pa X + i \al_0 \pa^2 X\quad,
}
where $2\al_0 = \lfr{p'-p}{\sqrt{pp'}}$.
The pure vertex operators of the form $V_\al = e^{i\al X}$ provide a
set of (Virasoro) primary fields of conformal dimension $\Delta_\al
=\al(\al - 2\al_0)$.
In this set, there are two operators of dimension
$1$ which correspond to the values of $\al$ given by
$\al_+\equiv\lfr{p'}{\sqrt{pp'}}$ and
$\al_-\equiv\lfr{-p}{\sqrt{pp'}}$. Consider the set of primary fields
labelled by $\al_{m',m}$ where $\al_{m',m}\equiv\lfr{(1-m')}2 \al_- +
\lfr{(1-m)}2 \al_+$ with $(m',m)$ restricted to the {\it conformal grid}
i.e., $0<m'<(p'-1)$ and $0<m<(p-1)$. It can be seen that the
dimensions of these operators agree with the Kac dimension formula for
the primary fields of the $(p',p)$ minimal model \footnote{$^a$}{Note that
the operators $V_\al$ and
$V_{(2\al_0-\al)}$ have the same conformal dimension.}. However, there is a
doubling -- this corresponds to the fact that given the dimension of
the primary field, there are two values of $\al$ which give the same
dimension. This is referred to as {\it minimal model duality}.
Hence, modulo this doubling one can obtain all the states of the
minimal models in this system. Dotsenko and Fateev have shown that in
order to reproduce the minimal model correlation functions with these
operators one has to introduce screening operators into the
correlation functions to saturate the background charge. The screening
operators are given by $Q_\pm \sim \int V_{\al_\pm}$. The screening
operators are of conformal dimension $0$ and hence their introduction
into correlation functions does not alter the conformal properties of
the correlator. On the sphere, the charge conservation equation
takes the form
\eqn\emcharge{
\sum_{i} \al_i + r~ \al_+ + r'~ \al_- = 2 \al_0
}
for a correlation function involving $V_{\al_i}$'s. We also have
introduced $r~Q_+$ and $r'~Q_-$ to obtain charge conservation.

Using the screening operator to obtain a BRST operator, Felder$^{\GF}$
has shown how the Hilbert space of a scalar field is truncated to the
finite set of primaries of the minimal models and their secondaries.
The BRST operator also explained the decoupling of
(Virasoro) null states in correlation functions.
All the primaries of the minimal model
correspond to states in the cohomology of this BRST operator(We shall
henceforth refer to this operator generically as $Q_F$). Further, all
null states are either $Q_F$-exact or not physical(i.e., not $Q_F$-closed).

Given a Virasoro secondary over a primary field, one can map it onto a
state in the Fock space using equation \estressa. This is done after
expanding the stress-tensor into modes. In general, this map is
non-vanishing. However, for some of the Virasoro nulls, this map
vanishes. We shall now illustrate this. Consider the identity operator
which is a conformal field of dimension $0$. Let us represent it by
$\ket{\Delta=0}$. It has a level one null -- $L_{-1}\ket{\Delta=0}$.
In the Fock space, due to minimal model duality, the identity is
represented by two different vertex operators: $V_{\al=0}$ and
$V_{\al=2\al_0}$. On mapping the level one null to the Fock space, we
have for
\eqn\enull{
\vbox{\baselineskip4pt\hbox{$L_{-1}\ket{\Delta=0}\mapsto
\left\{\phantom{\vrule height6mm}~~~\right.$}\hbox{}}
\vbox{\baselineskip12pt
\hbox{$ \pa~ (e^{2i\al_0 X}) =2i\al_0 \pa X~ e^{2i\al_0 X}$ for $\al=2\al_0$}
\hbox{}\hbox{ $ \pa~ 1 =0$ ~~~~~~~~~~~~~~~~~~~~~~~~~for $\al=0$ }
}\quad.
}
The vanishing of the null is clearly seen for $\al=0$. Since the
number of states over a primary at the given level are the same in the
Virasoro module as well as the Fock space(given by the number of
partitions of the level), there is  an oscillator state over
the primary at the same level which is not mapped onto from the
Virasoro module. So every {\it {vanishing null}} is replaced by an oscillator
state which we shall refer to as $\ket{w}$ or just `$w$'(=$\pa
X$ in the example just considered). The {\it non-vanishing
nul}l state will be referred to as $\ket{u}$ or just `$u$'(=$2i\al_0\pa
X e^{2i\al_0 X}$ in the example just considered). The $u$'s and $w$'s
will be useful in understanding minimal models after coupling to
gravity. On taking the dual in momenta, the nulls exchange their roles
i.e., $u\leftrightarrow w$.
\newsec{MINIMAL MODELS COUPLED TO GRAVITY}
We shall work in the conformal gauge and treat the Liouville mode
$\phi$ as a free field with (imaginary) background charge.
The stress-tensor for the Liouville field is
\eqn\estressb{
T_L =-\lfr14 \pa\phi \pa\phi + i \beta_0 \pa^2\phi\quad,
}
where $c_L=1-24 \beta_0^2$ and $c_M + c_L =26$. We  obtain that
for the $(p',p)$ model, $2\beta_0 = {{i(p+p')}\over{\sqrt{pp'}}}$.
The vertex operator $e^{i\beta\phi}$ has conformal weight
$\beta(\beta-2\beta_0)$.
Just as in the case of the critical bosonic string, in the conformal
gauge, physical states are in the cohomology of the string BRST. It is
given by
\eqn\ebrst{
Q_B = \oint :c(z) (T_M(z) + T_L(z) + \lfr12 T_{gh}(z)):\quad,
}
where $T_{gh}$ is the stress-tensor for the ghosts. The tensor product
of Fock spaces  $\CF(\al)\otimes\CF(\beta)\otimes\CF(gh)$ provides the
space on which $Q_B$ acts. Unlike, in the case
of the critical bosonic string one has to deal with the
cohomology of two BRST operators -- $Q_B$ and $Q_F$\footnote{$^b$}{Please
refer to  \mukhi\ for an introduction to $c<1$ non-critical
strings.}.

The physical
states of $c<1$ matter coupled to gravity has been studied in \LZ\ and
\BMP. They  have shown that there exist an infinite
number of BRST invariant states in $c<1$ theories coupled to gravity. The
Liouville momenta of these states are such as to provide the
gravitational dressing of the matter null states of the minimal model.
Further, there is one state at ghost number $\pm n$ for every matter Virasoro
representation whose Liouville
momenta are  $\beta > \beta_0$ for ghost number $+n$ states and $\beta <
\beta_0 $ for ghost number $-n$ states\footnote{$^c$}{The state $\gh$ is
assigned ghost number $0$. However, in going from states to operators,
we increase the ghost number by $1$. So the state $b_{-1}\gh$ has
ghost number $-1$ while the corresponding operator $b_{-1}c_1$ has
ghost number $0$.}. These states of non-trivial ghost number will be
called the LZ states.

\subsec{Descent Equations}
Due to the existence of two BRST operators,  there exist descent
equations which begin at  LZ states$^{\descent}$. It
was also shown in \descent\ that the descent equations relate LZ
states with pure vertex operators with matter momenta outside the
conformal grid. The simplest LZ state of ghost number $-1$ illustrates
descent equations.
\eqn\edesa{
Q_B\ket{LZ}^{-1}  = Q_F\ket{DK}^{0} =
\um{m'}{m}\otimes\ket\beta\otimes\gh \quad,
}
where $\um{m'}{m}$ is the `$u$'(non-vanishing null) over the Fock
primary  labelled by $(m',m)$. This generalises to
\eqn\edese{\eqalign{
Q_B\ket{LZ}^{-n}  &= Q_F\ket{I_1}^{-n+1}\quad,\cr
Q_B\ket{I_1}^{-n+1} &= -Q_F\ket{I_2}\quad,\cr
             &~\vdots \cr
Q_B\ket{I_{n-1}}^{-1} &=(-)^{n+1} Q_F\ket{DK}^{0}\quad,
}}
where $\ket{DK}^0=\dot{m'}{m}$ and the ghost-numbers are given by the
superscript. The matter labels $(m',m)$ take values from outside the
conformal grid with suitable gravitational dressing($\beta<\beta_0$).
These descent equations follow from
$(\ket{LZ}+\ket{I_1}+\ldots+\ket{DK})$ being  closed under
$(Q_B-(-)^GQ_F)$.
Dotsenko and Kitazawa$^{\DOT\KIT}$ used these states to calculate
three-point correlators and obtained agreement with matrix-model
results.

However, not all descents end at ghost number zero.
There are other descents which end at ghost number $-1$ states(or
ghost number zero as operators)$^{\ring}$. These descents begin with LZ states
with matter momentum dual to that in \edese.  Again, let us consider
the simpler example given in \edesa. On taking the dual in matter
momentum of the LZ state, the non-vanishing matter null is replaced by
the vanishing null(as explained earlier). Hence, we obtain
\eqn\edesc{
Q_B\ket{\widetilde{LZ}}^{-1}  = {\rm vanishing~~matter~~null} = 0\quad,
}
where the $\widetilde{LZ}$ refers to taking LZ state in \edesa\ with
its matter momentum flipped to its dual. For the general case, again a
vanishing null is encountered precisely one step earlier than in
\edese. The descent is
\eqn\edesd{\eqalign{
Q_B\ket{\widetilde{LZ}}^{-n}  &= Q_F\ket{I'_1}^{-n+1}\quad,\cr
Q_B\ket{I'_1}^{-n+1} &= -Q_F\ket{I'_2}\quad,\cr
             &~\vdots \cr
Q_B\ket{I'_{n-2}}^{-2} &=(-)^{n} Q_F\ket{R}^{-1}\quad,\cr
Q_B\ket{R}^{-1} &=0\quad.
}}

Hence, there are two
possible end-points for descents :
states at zero ghost number -- DK states and states
at ghost number $-1$. The latter are precisely the ring elements for $c<1$.
We shall describe them in the next subsection.

So far the discussion has been restricted to the negative ghost number
sector. The positive ghost number states are partners to the negative ghost
number states in the sense that the norm on the sphere is obtained as
\eqn\enorm{
\phantom{a}^{+n}\melt{LZ}{c_0}{LZ}^{-n}
}
Given an LZ state of ghost number $-n$, it is now trivial to construct a
$Q_B$ closed state of ghost number $+n$ which has a non-zero norm with
the given LZ state. This new state of ghost number $+n$ is
\eqn\epos{
|LZ\rangle^{+n}=M^n |\widehat{LZ}\rangle^{-n}\quad,
}
where $M=\{Q_B,c_0\}$ and $ |\widehat{LZ}\rangle^{-n}$ is the LZ state
with the matter and Liouville momenta flipped to their duals. The state
given in \epos\ is obviously not exact and hence a good element of the
cohomology. The Liouville dressing is $\beta>\beta_0$ as
given by the analysis of Lian and Zuckerman. It can also be shown that
the LZ states of positive ghost number thus obtained are equivalent to
those obtained by the construction described in \descent\  upto exact
pieces.

\subsec{Rings in $c<1$}
The general construction which argues for the presence of ghost number
$-1$ states in the Coulomb gas method allows us to develop a ring
structure in analogy with the work of Witten for $c=1$. Following the
suggestion of Kutasov, Martinec, and Seiberg\KMS, we define the two
operators that generate the (chiral)ring structure,
\eqn\egena{\eqalign{
x&=\CR_{1,2}=(b_{-2}c_1 + t(L_{-1}^{L} - L_{-1}^{M}))
e^{i\al_{1,2}X}e^{i\beta_{1,2}\phi}
\cr
y&=\CR_{2,1}=(b_{-2}c_1 + \lfr1t(L_{-1}^{L} - L_{-1}^{M}))
e^{i\al_{2,1}X}e^{i\beta_{2,1}\phi}
}}
i.e., $x$ and $y$ are the LZ$^{-1}$ states with matter momenta
labelled by $\al_{1,2}$ and $\al_{2,1}$ respectively and $t=\lfr p{p'}$.
We would like to point out that a target space  boost of the
$(X, \phi)$ system which transforms the $c=1$ theory to the appropriate
$c<1$ theory, would in fact transform the generators $x$ and $y$ of
Witten$^{\EW}$ to precisely the ones we have written above.
The full set of ring elements as well as DK states obtained depends on
the choice of screening operator to form the Felder BRST operator. We
refer to this as a choice of resolution. For the case of the unitary $(p+1,p)$
minimal models, the ring elements in the $Q_-$ resolution are$^{\ring}$
\eqn\eringminus{
(a_-)^n,~~ a_+(a_-)^n,~~ \ldots,~~(a_+)^{p-1}(a_-)^n\quad,
}
where $a_\pm$ are the non-chiral ring elements obtained by composing
holomorphic and anti-holomorphic ring elements -- $a_+ = -|x|^2$ and
$a_-=-|y|^2$. The DK states in the $Q_-$ resolution are
\eqn\edkminus{
V_n^\gamma = exp{{[p(n-2)+\gamma]\phi + [pn+\gamma+2]iX
 } \over {2\sqrt{p(p+1)}}}\quad,
}
where $\gamma=0,\ldots,(p-2)$ and $n=0,1,\ldots$. The value
$\gamma=(p-1)$ has been excluded since these Liouville exponents are
not seen in the (p+1)th critical point of p-matrix model$^{\BER}$.
They correspond to matter momenta of
the edge of the conformal grid with labels $(m',jp)$. However, their
decoupling from correlation functions is not obvious.

In the $Q_+$ resolution, the ring elements are given by
\eqn\eringplus{
(a_+)^n,~~ a_-(a_+)^n,~~ \ldots,~~(a_-)^{p}(a_+)^n\quad.
}
How are the ring elements in \eringminus\ and \eringplus\ related? An
equivalence relation $a_-^{p+1}\sim a_+^{p}$ imposed on the ring of
monomials of the form $\left \{a_+^m a_-^n\right \}$ makes the two
rings isomorphic to each other. The DK states in the $Q_+$ resolution are
\eqn\edkplus{
V_n^\gamma = exp{{[(p+1)(n-2)+\gamma+2]\phi + [-(p+1)n-\gamma]iX}
 \over {2\sqrt{p(p+1)}}}
}
where $\gamma=0,\ldots,(p-1)$ and $n=0,1,\ldots$. Here we have
excluded edges of the type $(j(p+1),m)$ which correspond to
$\gamma=p$. The Liouville exponents belong to those seen in the pth
critical point of the (p+1) matrix model$^{\BER}$.
\subsec{Correlation Functions}
Given a resolution, one would like to calculate correlation functions
involving arbitrary numbers of DK states. This has been accomplished
recently$^{\corr}$. The ring elements in \eringminus\ and \eringplus\
were utilised to obtain recursion relations amongst the DK states.
These recursion relations are useful in converting integrals not in
the form given by Dotsenko and Fateev into those which are of that
form. A {\it complete match} with matrix model results has been obtained
for all correlation functions on the sphere albeit with analytic
continuation  in the numbers of the cosmological constant operators
and matter screening operators. See ref. \corr\ for more details.
This completes the identification of matrix model observables as DK
states in continuum Liouville theory.

However, one would like to see how the matrix model observables can be
represented  not as pure vertex operator states but rather as states
built over the primaries of the minimal models. A suggestion has been
presented in ref. \ring. The states which have been presented as
candidates are given by
\eqn\ematrix{
M^{n}~~\ket{LZ}^{-n}\otimes \ket{\overline{LZ}}^{-n}\quad,
}
where $M\equiv \{Q_B + \bar{Q}_B, (c_0 - \bar{c}_0) \}$. Not only do
these states have total ghost number zero but also have the required
scaling dimensions. One can now easily construct ghost number conserving
correlation functions with these states. However, calculations involving
these states will need techniques along the line used in the context of
the string theory dilaton$^{\TCCT}$. However, one can hope to relate
this to the pure vertex operator states(DK states) using the descent
equation in correlation functions as described in ref. \descent.

\noindent {\bf Acknowledgements:}
I would like to thank T. Jayaraman and V. John for a collaboration
which has led to the work described here. I
would also like to thank the organisers of the workshop for an
opportunity to present this work at the Workshop.
\bigskip
\listrefs
\bye